\documentclass[summary]{emts2025}

\usepackage{subfig}
\usepackage{siunitx}
\usepackage{upgreek}
\usepackage{newtxmath}

\renewcommand{\vec}[1]{\boldsymbol{#1}}
\newcommand{\dyad}[1]{\boldsymbol{\bar{#1}}}

\usepackage{atbegshi}
\makeatletter
\def\ps@firstpage{%
\def\@oddfoot{\mycopyrightnotice}%
\def\@oddhead{}%
\def\@evenhead{}%
\def\@evenfoot{}%
}
\AtBeginShipout{\AtBeginShipoutUpperLeft{%
\put(\dimexpr\paperwidth-1cm\relax,-.55cm){\makebox[0pt][r]{
\begin{minipage}{\textwidth}
\centering \scriptsize
This article has been published in the Proceedings of the 2025 International Symposium on Electromagnetic Theory (URSI EMTS 2025). © 2025 IEEE. Personal use of this material is permitted. Permission from IEEE must be obtained for all other uses, in any current or future media.
Citation information: DOI: 10.46620/URSIEMTS25/RKHF8213
\end{minipage}
}}}%
}
\def\mycopyrightnotice{%
\begin{minipage}{\textwidth}
\centering \scriptsize
Copyright~\copyright~2025 URSI. Personal use of this material is permitted. Permission from URSI must be obtained for all other uses, in any current or future media, including\\reprinting/republishing this material for advertising or promotional purposes, creating new collective works, for resale or redistribution to servers or lists, or reuse of any copyrighted component of this work in other works.
\end{minipage}
}
\makeatother

\title{Polarization-Aware Ray-Tracing Enhanced Back-Projection Algorithm for Microwave Imaging in Complex Multipath Environments}

\author{Han Na\affref{ref1}, Quanfeng Wang\affref{ref1}, Matthias Saurer\affref{ref1}, Meisong Tong\affref{ref2}, and Thomas F. Eibert\affref{ref1}}

\affiliation{%
  \aff{ref1}{Department of Electrical Engineering, School of Computation, Information and Technology, \\ Technical University of Munich, Munich, Germany, han.na@tum.de}
  \aff{ref2}{Department of Electronic Science and Technology, Tongji University, Shanghai, China}
}

\begin{document}

\maketitle
\thispagestyle{firstpage}

\begin{abstract}
	A ray-tracing (RT) enhanced back-projection algorithm (RT-BPA) for microwave imaging in multipath environments is presented. By tightly incorporating the concept of ray-tracing into a generalized version of traditional BPA, this method ensures improved image quality by addressing two important issues. First, when the line-of-sight (LOS) path is obstructed, reflected paths, if available, enable imaging of hidden targets, which extends the applicability of the standard BPA beyond its normal use case. Second, the consideration of reflected ray-paths is equivalent to virtually increasing the aperture size, thus, improving image resolution without requiring new measurements. A key factor in achieving these advancements is the consideration of the vector nature of electromagnetic waves with polarization-dependent phase compensation, which is often ignored when employing a scalar-wave based formulation of the electromagnetic vector field. In addition, the presented method employs a shooting and bouncing rays (SBR) framework, offering better flexibility compared to manual path evaluation in existing  RT-BPAs.
\end{abstract}

\section{Introduction}
In microwave imaging, multipath effects caused by environmental scattering can be both beneficial and challenging. According to existing literature, these effects can be viewed from two perspectives. One possibility is to treat reflections from objects other than the target of interest as noise or clutter that requires suppression \cite{Leigsnering.2014}. Alternatively, such reflections can be interpreted and utilized  as indirect viewing mechanisms, enabling imaging when the direct line-of-sight (LOS) path is blocked \cite{Setlur.2014} or improving image resolution by effectively increasing the synthetic aperture \cite{Zhou.2015}. In this work, we focus on the latter perspective and introduce a ray-tracing (RT) enhanced back-projection algorithm (RT-BPA) for microwave imaging in multipath environments.

To address the challenges posed by multipath propagations, various studies have employed ray-based techniques to identify multipath contributions and correct signal phase based on ray-path lengths \cite{Zhang.2023, Rhiem.2024}. However, most of these approaches focus on the signal processing perspective, neglecting critical physical properties such as the polarization of electromagnetic waves. Without proper consideration of polarization effects, the coherent combination of various signal contributions from multiple paths becomes inaccurate. Thus, the quality of reconstructed images is reduced. Furthermore, many of these methods rely on fixed or simplified environments, explicitly calculating ray-paths for specific cases \cite{Zhang.2023} and often restricting the analysis to 2-D scenarios \cite{Setlur.2014}. Advanced imaging methods based on solving the inverse electromagnetic problem can theoretically address these limitations, but their reliance on analytically known Green's functions \cite{Saurer.2022, Wang.2025} makes them impractical for non-canonical or complex propagation environments.

This work introduces an enhancement to the standard formulation of the naive BPA using a general RT engine based on the CUDA OptiX framework~\cite{Parker.2010}. By employing the shooting and bouncing rays (SBR) method, the algorithm dynamically identifies ray-paths, enabling simulations of complex 3-D environments without manual path calculations. Employing a polarization-aware description of the scattering process, the proposed method improves directional resolution and serves as a valuable tool for creating well-focused microwave images in complex propagation environments.


\section{Theoretical Background}
Microwave imaging algorithms, including BPA, are commonly based on a linearization of the forward scattering problem according to the first-order Born~\cite{Saurer.2022,Saurer.Jan.2025}, which neglects mutual coupling between different points of the target scatterer. Due to this linearization, the forward scattering operator is expressed as an integral operator connecting the dyadic scattering distribution \( \dyad{s}_{\mathrm{B}}(\vec{r}) \) within the volume \( V_{\mathrm{S}} \) to the measurement samples \( T(\vec{r}_{\mathrm{T}},\vec{r}_{\mathrm{R}},k) \) according to~\cite{Saurer.Jan.2025}
\begin{equation}\label{eq:forward_op}
T(\vec{r}_{\mathrm{T}},\vec{r}_{\mathrm{R}},k)=
		\iiint\limits_{V_{\mathrm{S}}}
		\vec{E}_{\mathrm{R}}\left({\vec{r}},\vec{r}_{\mathrm{R}}\right)
		\cdot
		\dyad{s}_{\mathrm{B}}\left({\vec{r}}\right)
		\cdot
		\vec{E}_{\mathrm{T}}\left({\vec{r},\vec{r}_{\mathrm{T}}}\right)
		\,
		\mathrm{d}^3 \vec{r}\,,
\end{equation}
where $k$ is the wavenumber while $\vec{r}_{\mathrm{T}}$ and $\vec{r}_{\mathrm{R}}$ are the position vectors of the transmitter (Tx) and receiver (Rx), respectively. Additionally,  the expression $\vec{E}_{\mathrm{T}}\left({\vec{r},\vec{r}_{\mathrm{T}}}\right)$ in (\ref{eq:forward_op})
characterizes the electric field vector of the Tx antenna, which is incident on the scattering object, while $\vec{E}_{\mathrm{R}}\left({\vec{r},\vec{r}_{\mathrm{R}}}\right)$ is the scattered field vector, which is reciprocal to the field radiated by the Rx antenna, when it is used in transmit mode~\cite{Saurer.Jan.2025}. 

In the standard BPA, the wave propagation environment between each Tx and the scatterer, as well as between the scatterer and the Rx, is assumed to be free space~\cite{Gumbmann.2017}.
Denoting polarization vectors $\vec{\hat{E}}_{\mathrm{T}}(\vec{r},\vec{r}_{\mathrm{T}})$ and $\vec{\hat{E}}_{\mathrm{R}}(\vec{r},\vec{r}_{\mathrm{R}})$, respectively, and 
decomposing the field propagation operator into a linear phase shift term $\exp\left[-\mathrm{j}\varphi(\vec{r})\right]$ and a magnitude term $A(\vec{r}_{\mathrm{T}},\vec{r}_{\mathrm{R}},\vec{r})$\footnote{In this work, the attenuation factors are set to 1 for arbitrary $\vec{r}_{\mathrm{T}}$, $\vec{r}_{\mathrm{R}}$ and $\vec{r}$ by employing the paraxial approximation.}, (\ref{eq:forward_op}) can be rewritten as
\begin{multline}\label{eq:forw_op_vec}	T(\vec{r}_{\mathrm{T}},\vec{r}_{\mathrm{R}},k)=
		\iiint\limits_{V_{\mathrm{S}}}
\vec{\hat{E}}_{\mathrm{R}}(\vec{r},\vec{r}_{\mathrm{R}})\cdot\dyad{s}_{\mathrm{B}}\left({\vec{r}}\right)\cdot\vec{\hat{E}}_{\mathrm{T}}(\vec{r},\vec{r}_{\mathrm{T}})\\
A(\vec{r}_{\mathrm{T}},\vec{r}_{\mathrm{R}},\vec{r})\mathrm{e}^{-\mathrm{j}\varphi(\vec{r})} 
		\mathrm{d}^3 \vec{r}\,,
\end{multline}
where 
$\varphi(\vec{r})=k\left[R_\text{Rx}(\boldsymbol{r}) + R_\text{Tx}(\boldsymbol{r})\right]$
with
 $R_\text{Tx}(\boldsymbol{r}) =\left|\vec{r}-\vec{r}_{\mathrm{T}}\right|$ and $R_\text{Rx}(\boldsymbol{r}) =\left|\vec{r}-\vec{r}_{\mathrm{R}}\right|$ denoting the corresponding path lengths between the antennas and a specific point located in the imaging domain. However, this formulation neglects the wave interaction with the background objects, e.g., walls and ceilings in an indoor scenario~\cite{Leigsnering.2014}.
 With the consideration of reflective propagation environments,
 we separate the field vectors interacting with the scattering object into a superposition of distinct wavefronts characterized by a unique index $w$
 , and (\ref{eq:forw_op_vec}) is generalized to
\begin{multline}\label{eq:forw_op_gen}	T(\vec{r}_{\mathrm{T}},\vec{r}_{\mathrm{R}},k)=
		\iiint\limits_{V_{\mathrm{S}}}\sum\limits_w	\vec{\hat{E}}^w_{\mathrm{R}}(\vec{r},\vec{r}_{\mathrm{R}})\cdot\dyad{s}_{\mathrm{B}}\left({\vec{r}}\right)\cdot\vec{\hat{E}}^w_{\mathrm{T}}(\vec{r},\vec{r}_{\mathrm{T}})\\A^w(\vec{r}_{\mathrm{T}},\vec{r}_{\mathrm{R}},\vec{r})\mathrm{e}^{-\mathrm{j}\varphi^w(\vec{r})} 
		\mathrm{d}^3 \vec{r}\,.
\end{multline}
By taking the adjoint of the generalized forward scattering operator in multipath environments, as given in (\ref{eq:forw_op_gen}), and considering only one set of co-polarized scattering data for a fair comparison with the conventional BPA, we get
\begin{multline}\label{eq:adjoint_op_gen}
    s_{\mathrm{B}}(\vec{r})=\sum\limits_{w}\int\limits_0^{\infty}\iint\limits_{L_{\mathrm{T}}}\iint\limits_{L_{\mathrm{R}}}  A^w(\vec{r}_{\mathrm{T}},\vec{r}_{\mathrm{R}},\vec{r})T(\vec{r}_{\mathrm{T}},\vec{r}_\mathrm{R},k)\\\mathrm{e}^{\,\mathrm{j}\varphi^w(\vec{r})+\mathrm{j}\uppi\delta(w)}\mathrm{d}^2\vec{r}_{\mathrm{R}}\,\mathrm{d}^2\vec{r}_{\mathrm{T}}\,\mathrm{d}k\,,
\end{multline}
where \( L_{\mathrm{T}} \) and \( L_{\mathrm{R}} \) represent
the aperture of the transmit and receive array, respectively.
The term \( \delta(w) \) is either \( 1 \) for parallel polarization or \( 0 \) for vertical polarization.
This formulation accounts for a \( \uppi \)-phase shift due to potential polarization changes, a phenomenon also known as the half-wave loss in optics \cite{Born.1999}.
By comparison, the reconstruction formula for the naive BPA is given by
\begin{equation}\label{eq:naive_BPA}
    s_{\mathrm{B}}(\vec{r}')=\int\limits_0^{\infty}\iint\limits_{L_{\mathrm{T}}}\iint\limits_{L_{\mathrm{R}}}T(\vec{r}_{\mathrm{T}},\vec{r}_\mathrm{R},k)\,\mathrm{e}^{\,\mathrm{j}\varphi(\vec{r})} \mathrm{d}^2\vec{r}_{\mathrm{R}}\,\mathrm{d}^2\vec{r}_{\mathrm{T}}\,\mathrm{d}k,
\end{equation}
which does not account for multipath effects.
By incorporating the separation of different wavefronts and a polarization-specific phase-compensation, the RT-BPA, expressed in (\ref{eq:adjoint_op_gen}), is expected to achieve improved focusing capabilities compared to the naive BPA in (\ref{eq:naive_BPA}). With slight modifications, e.g., dropping the integral over the receive array, (\ref{eq:adjoint_op_gen}) and (\ref{eq:naive_BPA}) can also be utilized in radiation problems.

With the proper consideration of different wavefronts, the generalized BPA expressed by (\ref{eq:adjoint_op_gen}) can be easily incorporated with ray-tracing frameworks using the SBR approach.
Here, rays are randomly launched from each image voxel and propagate through a 3-D environment modeled by a triangular mesh following geometrical optics (GO) principles.
Each ray captured by a Tx or an Rx provides the optical path length information and determines whether an extra 
$\uppi$-phase shift is introduced in (\ref{eq:adjoint_op_gen}).
The latter is managed by assigning an integer hash identifier $w$ to each individual path, which depends on the sequence of the environmental objects it interacts with. 
Since all rays experiencing the same bounces share a common identifier, $w$, the post-processing of the extra phase shift denoted by $\delta(w)$ in (\ref{eq:adjoint_op_gen}) is easy and efficient.
The final image is then created by coherently summing up all signal contributions from the obtained wavefronts.
To manage the computational demands, the CUDA OptiX framework is utilized for parallelized ray-tracing, along with path prediction techniques \cite{Na.2023} to reduce the number of rays needed.



\section{Numerical Results and Discussion}
\textit{A. \textbf{Imaging Hidden Objects without LOS-paths}}

\begin{figure}
	\centering
	\subfloat[]{\includegraphics[scale=0.65]{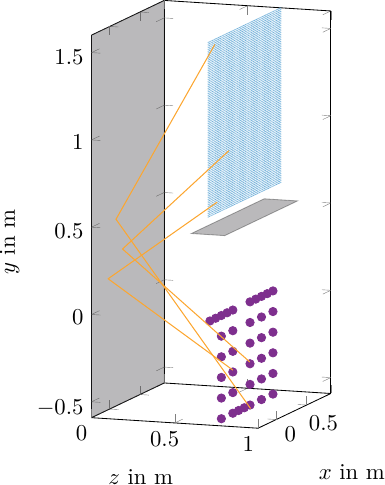}}
	\subfloat[]{\includegraphics[scale=0.65]{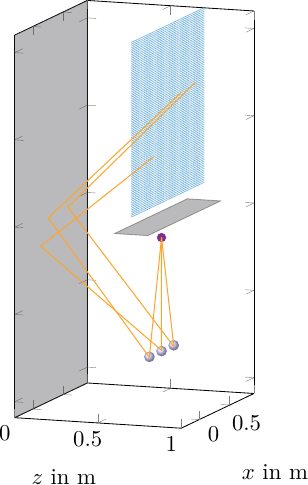}}
	\caption{(a) Radiaton problem including several Hertzian dipoles forming a `TUM' logo in front of an infinitely large ground plane. (b) Scattering problem consisting of three PEC spheres illuminated by a single dipole source. In both cases the line of sight is blocked by a thin sheet of PEC.} 
	\label{vis_sim}
\end{figure}

To assess the effectiveness of the proposed RT-BPA in imaging hidden objects, two different simulations, as depicted in Figure~\ref{vis_sim}, are conducted using the commercial full-wave simulation software FEKO~\cite{EMSS2025}. The simulation frequency in both cases ranges from $\SI{18}{\giga\hertz}$ to $\SI{20}{\giga\hertz}$ with an equal frequency step of $\SI{100}{\mega\hertz}$. In both scenarios, an infinitely large perfect electrically conducting (PEC) ground plane is located at $z=\SI{0}{\meter}$, and the LOS is blocked by another thin PEC plate.

\begin{figure}
	\centering
	\subfloat[]{\includegraphics[scale=0.58]{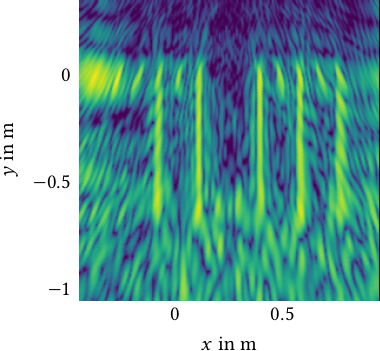}}\hfill
	\subfloat[]{\includegraphics[scale=0.58]{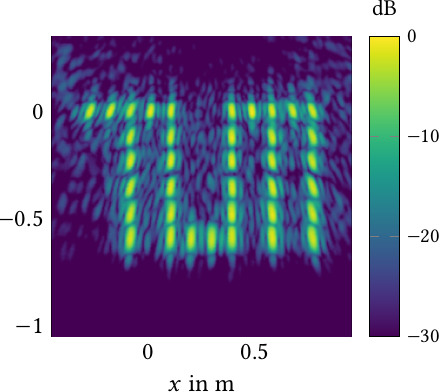}}
	\caption{Reconstruction results for the dipole radiators forming a `TUM' logo according to Figure \ref{vis_sim}(a) while utilizing the standard BPA with only LOS contributions (a) and the RT-BPA (b).}
        \label{TUM_images}
\end{figure}

The first scenario consists of 37 Hertzian dipoles arranged to represent a `TUM' logo. The dipoles are radiating at a distance of $\SI{0.7}{\meter}$ above the PEC ground plane. One co-polarized component of the electric field, which was computed in FEKO at a total of $\num{10000}$ regularly distributed observation points, spanning an area of $\SI{1.2}{\meter}$ by $\SI{1}{\meter}$, serve as the input data for the two imaging methods. The reconstruction result for the naive BPA, shown in Figure~\ref{TUM_images}(a), is highly distorted and allows only for a rough identification and localization of the dipole radiators.
In contrast, the result for the RT-BPA, as shown in Figure~\ref{TUM_images}(b), shows significantly better focusing of the dipoles.
This is because our method correctly accounts for the phase shift of the waves when they propagate from the sources toward the observation locations, including the reflection at the ground plane. 

\begin{figure}
	\centering
	\subfloat[]{\includegraphics[scale=0.58]{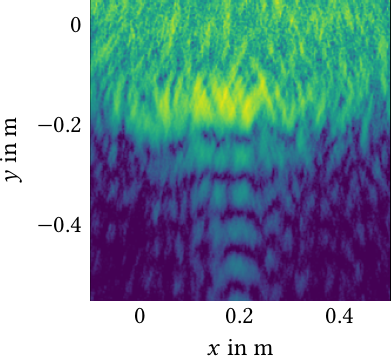}}\hfill
	\subfloat[]{\includegraphics[scale=0.58]{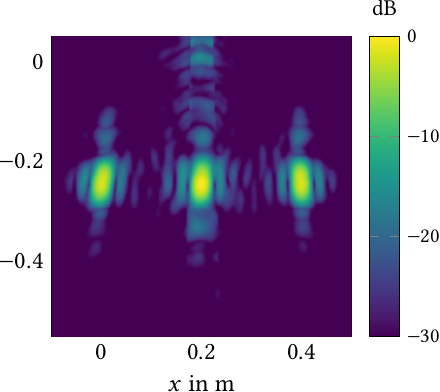}}
	\caption{Reconstruction results for the scattering of three PEC spheres as given in Figure \ref{vis_sim}(b) while  employing the standard BPA with only LOS contributions (a) and the RT-BPA (b).} 
        \label{Scattering_images}
\end{figure}
In the second case, we slightly modify the previously discussed setup by considering only one single Hertzian dipole positioned at coordinates $\left(x,y,z\right)=\left(0.2, 0.4, 0.7\right)\,$m. In this case, we are interested in the scattering of three small PEC spheres with a radius of $\SI{0.02}{\meter}$,  which are arranged in a line along the $x$-direction at coordinates $\left(0, -0.25, 0.7\right)\,$m, $\left(0.2, -0.25, 0.7\right)\,$m and $\left(0.4, -0.25, 0.7\right)\,$m, respectively. The spatial images for this scattering scenario are given in Figure~\ref{Scattering_images}. In this case, the reconstruction result for the naive BPA only assuming LOS contributions is uninterpretable. In contrast, the image for our RT-BPA given in Figure~\ref{Scattering_images}(b) is still capable of computing a well-focused image, where all three PEC spheres are clearly recognizable. 

\textit{B. \textbf{Improving Image Resolution by Using Higher-Order Reflection Paths }}

\begin{figure}
	\centering
	\includegraphics[scale=0.4]{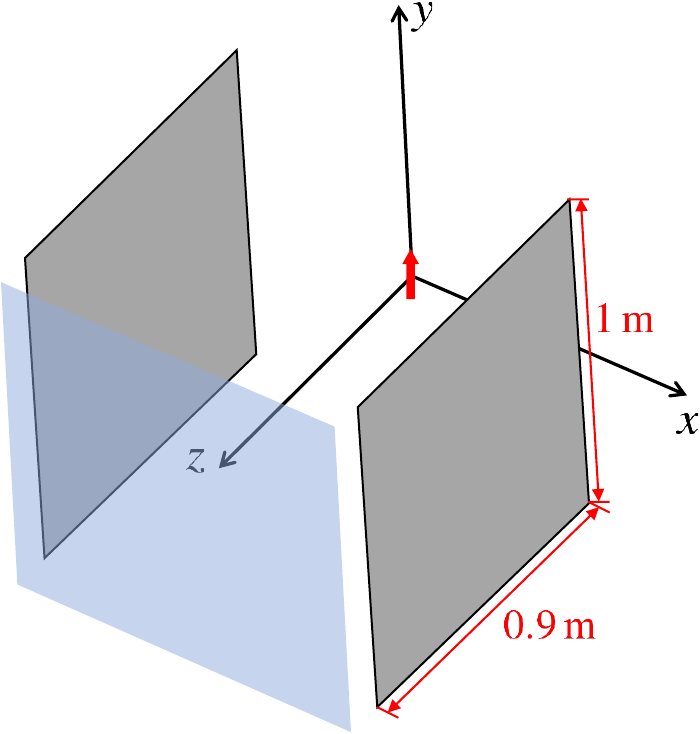}
	\caption{Illustration of the simulation setup featuring a single dipole antenna positioned between two parallel PEC plates.}
	\label{fig:plate2}
\end{figure}

\begin{figure}
  \centering
  \subfloat[]{\includegraphics[scale=0.56]{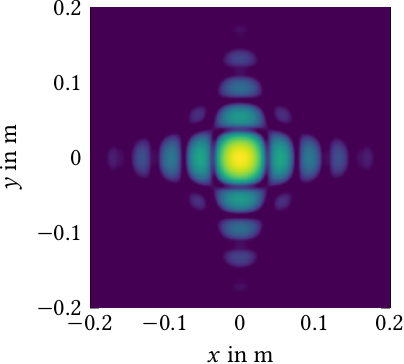}}%
  \hfill
  \subfloat[]{\includegraphics[scale=0.56]{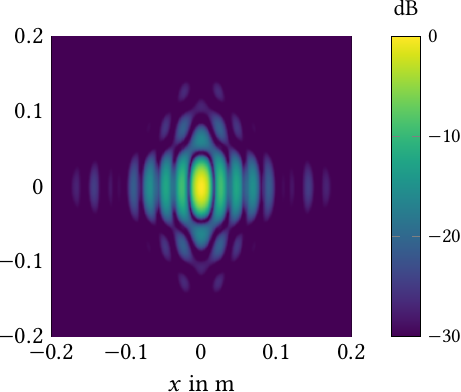}}%
  \\ 
  \vspace{-5mm}
  \subfloat[]{\includegraphics[scale=0.56]{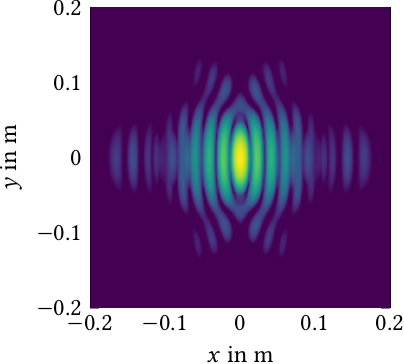}}%
   \hfill
  \subfloat[]{\includegraphics[scale=0.56]{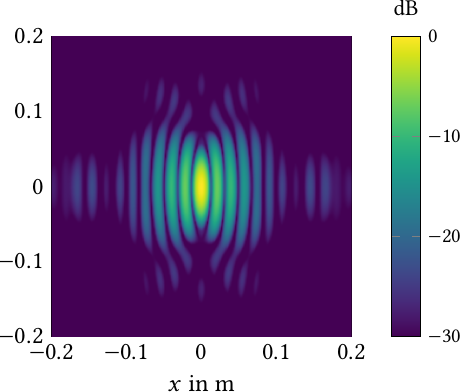}}%
  \caption{Imaging results of the dipole using standard BPA with LOS contributions~(a), and RT-BPA incorporating reflection contributions up to first-order~(b), second-order~(c), and third-order~(d).}
  \label{fig:resolution}
\end{figure}

In order to validate that the utilization of multipath contributions can also enhance imaging quality, the simulation setup illustrated in Figure~\ref{fig:plate2} is considered next, where a single Hertzian dipole acting as the radiation source is placed between two parallel PEC plates. The signal contributions originating from both the LOS and 
 multiple reflections within the plates are captured by a set of co-polarized Hertzian dipoles located in the planar measurement surface as shown in Figure~\ref{fig:plate2}.  The imaging result for the standard BPA, which  only accounts for LOS contributions and discards multiple reflections, is displayed in Figure~\ref{fig:resolution}(a). By comparison, the imaging results obtained by the RT-BPA incorporating up to first, second, and third-order reflections are shown in Figures~(b), (c), and (d), respectively. It is seen that the utilization of higher-order reflections significantly improves the imaging resolution along the $x$-dimension, which is equivalent to a virtually enlarged measurement plane along that direction. Utilizing 256 by 256 pixels as well as an NVIDIA RTX A6000 GPU, the proposed RT-BPA achieved for all cases run times smaller than 20 seconds, which is comparable to the reconstruction times for the non-LOS scenarios from the previous example.

\section{Conclusion}
A polarization-aware ray-tracing enhanced back-projection algorithm for microwave imaging in complex multipath environments was presented and validated with various numerical examples. Utilizing an SBR framework, the propagation paths between a set of transmitting and receiving antennas were separated into different wavefronts, enabling improved focusing capabilities compared to standard back-projection algorithms. 
Additionally, our CUDA-OptiX-based implementation demonstrated high computational efficiency due to massive parallelization and sophisticated path prediction strategies.


\section{Acknowledgements}
Funded by the European Union. Views and opinions expressed
are however those of the author(s) only and do not
necessarily reflect those of the European Union or European
Innovation Council and SMEs Executive Agency (EISMEA).
Neither the European Union nor the granting authority can be
held responsible for them. Grant Agreement No: 101099491.

\bibliographystyle{IEEEtran}
\bibliography{literature.bib}

\end{document}